\documentclass[twocolumn,showpacs,aps,prd]{revtex4}
\usepackage{graphicx}
\usepackage{dcolumn}
\usepackage{amsmath}
\usepackage{epsfig}
\usepackage{hyphenat}
\long\def\inst#1{\par\nobreak\kern 4pt\nobreak
    {\itshape #1}\par\vskip 10pt plus 3pt minus 3pt}
\RequirePackage{xspace}
\usepackage{relsize}

\def\babar{\mbox{\slshape B\kern-0.1em{\smaller A}\kern-0.1em
    B\kern-0.1em{\smaller A\kern-0.2em R}}}
\def\Kbar    {\kern 0.18em\overline{\kern -0.18em K}{}\xspace}

\def\Kz      {\ensuremath{K^0}\xspace}
\def\Kzb     {\ensuremath{\Kbar^0}\xspace}
\def\KzKzb   {\ensuremath{\Kz {\kern -0.16em \Kzb}}\xspace}

\def\Ks     {\ensuremath{K_S}\xspace}
\def\Kl     {\ensuremath{K_L}\xspace}
\def\KsKs   {\ensuremath{\Ks {\kern -0.16em \Ks}}\xspace}
\def\KlKl   {\ensuremath{\Kl {\kern -0.16em \Kl}}\xspace}
\def\KsKl   {\ensuremath{\Ks {\kern -0.16em \Kl}}\xspace}
\def\KlKs   {\ensuremath{\Kl {\kern -0.16em \Ks}}\xspace}
\def\Dbar    {\kern 0.18em\overline{\kern -0.18em D}{}\xspace}
\def\Dz      {\ensuremath{D^0}\xspace}
\def\Dzb     {\ensuremath{\Dbar^0}\xspace}
\def\DzDzb   {\ensuremath{\Dz {\kern -0.16em \Dzb}}\xspace}
\def\Bbar    {\kern 0.18em\overline{\kern -0.18em B}{}\xspace}

\def\Bz      {\ensuremath{B^0}\xspace}
\def\Bzb     {\ensuremath{\Bbar^0}\xspace}
\def\BzBzb   {\ensuremath{\Bz {\kern -0.16em \Bzb}}\xspace}
\def\Bu      {\ensuremath{B^+}\xspace}
\def\Bub     {\ensuremath{B^-}\xspace}

\def\BpBm    {\ensuremath{\Bu {\kern -0.16em \Bub}}\xspace}

\newcommand{\optbar}[1]{\shortstack{{\tiny (\rule[.4ex]{1em}{.1mm})}
  \\ [-.7ex] $#1$}}
\def\BorBbar    {\kern 0.18em\optbar{\kern -0.18em B}{}\xspace}
\def\DorDbar    {\kern 0.18em\optbar{\kern -0.18em D}{}\xspace}
\def\KorKbar    {\kern 0.18em\optbar{\kern -0.18em K}{}\xspace}

\def\pep2{PEP-II}
\mathchardef\Upsilon="7107
\def\Y#1S{\ensuremath{\Upsilon{(#1S)}}\xspace}% no space before {...}!

\begin{document}

\title{
\large \bfseries \boldmath Search for $\eta$ and $\eta'$ Invisible Decays in $J/\psi\to\phi\eta$ and $\phi\eta'$}
\author{\small
M.~Ablikim$^{1}$, M.~N.~Achasov$^{6}$, O.~Albayrak$^{3}$, D.~J.~Ambrose$^{39}$, F.~F.~An$^{1}$, Q.~An$^{40}$, J.~Z.~Bai$^{1}$, Y.~Ban$^{26}$, J.~Becker$^{2}$, J.~V.~Bennett$^{16}$
% ...
\\
\vspace{0.2cm}
(BESIII Collaboration)\\
\vspace{0.2cm} {\it
$^{1}$ Institute of High Energy Physics, Beijing 100049, People's Republic of China\\
$^{2}$ Bochum Ruhr-University, D-44780 Bochum, Germany\\
$^{3}$ Carnegie Mellon University, Pittsburgh, Pennsylvania 15213, USA\\
% ...
$^{6}$ G.I. Budker Institute of Nuclear Physics SB RAS (BINP), Novosibirsk 630090, Russia\\
% ...
$^{16}$ Indiana University, Bloomington, Indiana 47405, USA\\
% ...
$^{26}$ Peking University, Beijing 100871, People's Republic of China\\
% ...
$^{39}$ University of Rochester, Rochester, New York 14627, USA\\
$^{40}$ University of Science and Technology of China, Hefei 230026, People's Republic of China\\
% ...
}}

%%%%%%%%%%%%%%%%%%%%%%%%%%%%%%%%%%%%%%%%%%%%%%%%%%%%%%%%%%%%%%%%%%%%%%%%%%%%%%%%%%
%                             DATE                                              %%
%%%%%%%%%%%%%%%%%%%%%%%%%%%%%%%%%%%%%%%%%%%%%%%%%%%%%%%%%%%%%%%%%%%%%%%%%%%%%%%%%%

%\date{\today}
%\date{March 32, 2003}

%%%%%%%%%%%%%%%%%%%%%%%%%%%%%%%%%%%%%%%%%%%%%%%%%%%%%%%%%%%%%%%%%%%%%%%%%%%%%%%%%%
%                             Abstract                                          %%
%%%%%%%%%%%%%%%%%%%%%%%%%%%%%%%%%%%%%%%%%%%%%%%%%%%%%%%%%%%%%%%%%%%%%%%%%%%%%%%%%%

\begin{abstract}

Using a sample of $(225.3\pm 2.8)\times 10^{6}$ $J/\psi$ decays
collected with the BESIII detector at BEPCII, searches for invisible
decays of $\eta$ and $\eta^\prime$ in $J/\psi\to\phi\eta$ and
$\phi\eta^\prime$ are performed. Decays of $\phi \to K^{+}K^{-}$ are
used to tag the $\eta$ and $\eta^\prime$ decays. No signals above
background are found for the invisible decays, and upper limits at
the $90\%$ confidence level are determined to be $2.6\times10^{-4}$
for the ratio
$\frac{\mathcal{B}(\eta\to\rm{invisible})}{\mathcal{B}(\eta\to\gamma\gamma)}$
and $2.4\times10^{-2}$ for
$\frac{\mathcal{B}(\eta^\prime\to\rm{invisible})}{\mathcal{B}(\eta^\prime
\to\gamma\gamma)}$. These limits may be used to constrain light dark
matter particles or spin-1 $U$ bosons.

\end{abstract}

% PACS, the Physics and Astronomy Classification Scheme.

\maketitle

%\tolerance=1
%\emergencystretch=\maxdimen
%\hyphenpenalty=10000
%\hbadness=10000

%%%%%%%%%%%%%%%%%%%%%%%%%%%%%%%%%%%%%%%%%%%%%%%%%%%%%%%%%%%%%%%%%%%%%%%%%
% INTRODUCTION
%%%%%%%%%%%%%%%%%%%%%%%%%%%%%%%%%%%%%%%%%%%%%%%%%%%%%%%%%%%%%%%%%%%%%%%%%
\section{Introduction}
\label{sec:intro} Invisible or radiative decays of the
$J/\psi,\,\Upsilon$ and other mesons may be used to search for new
physics beyond the Standard Model (SM), in particular for neutral
states $\chi$, that could be light dark matter constituents,
according to $\,q\bar q\to \,(\gamma)\,\chi\chi$
~\cite{np0,np1,np2bis}. Independently of dark matter, radiative
meson decays into $\gamma$ + invisible allow to look, as for
\hbox{spin-0} axions\,\cite{ax}, for light spin-1 particles called
$U$ bosons, according to $\,q\bar q\to\gamma + U $, where the $U$
can stay invisible when decaying into $\,\nu\bar\nu$ or other
neutral particles~\cite{npa,prd75}. Such $J/\psi$ or $\Upsilon \to
\gamma + U$ decays were already searched for long
ago~\cite{edwards,balest,JInsler}.

Processes involving $U$ bosons and dark matter particles $\chi$ may
be intimately related, with the $U$'s mediating a new interaction
between ordinary (SM) and dark matter particles $\chi$. This  may
indeed be necessary to ensure for sufficient annihilations of {\it
light} dark matter (LDM) particles\,\cite{bf}, proposed as an
interpretation for the \nohyphens{origin} of the 511 keV line from
the galactic bulge observed by the INTEGRAL
satellite~\cite{np3,np8}.

Conversely, this interaction mediated by $U$ bosons may be
responsible for the pair-production of LDM particles through $q\bar
q$ (or $e^+e^-$) $\to (\gamma) \,\chi\chi$. In spite of tentative
estimates like $\mathcal{B}(\eta\, (\eta^\prime) \rightarrow \chi
\chi)\approx 1.4\times 10^{-4}~(1.5\times 10^{-6}$)~\cite{np10}, one
cannot reliably predict such invisible decay rates of mesons just
from the dark matter relic density and annihilation
cross-sec\-tion~\cite{np2bis}. In particular a $U\!$ vectorially
coupled to quarks and leptons could be responsible for LDM
annihilations, without contributing to invisible decays
$\eta\,(\eta')\,\to\chi\chi$~\cite{np1}; this includes the more
specific case of a  $U$ boson coupled to SM particles through the
electromagnetic current~\cite{npb90}, also known as a ``dark
photon''. Annihilations $q\bar q\to UU$ may also be a source of
invisible meson decays, especially as the invisible decay mode $U
\to \chi\chi$ may be dominant~\cite{np1}. $U$ exchanges could be
responsible for a possible discrepancy between the measured and
expected values of $g_\mu-2$~\cite{prd75}.

It is in any case very interesting to search for such light
invisible particles in collider experiments~\cite{np3-0}. Many
searches for the invisible decays of
$\pi^0,\,\eta,\,\eta^\prime,\,J/\psi$ and $\Upsilon(1S)$ have been
performed~\cite{np4,np5,cleo-c,np6,np7}. Invisible decays of $\eta$
and $\eta'$ may originate from $\eta \,(\eta')\,\to \chi\chi$ or
$U_{\rm inv} U_{\rm inv}$. The resulting informations complement
those from $\,J/\psi\,$ and $\Upsilon$ decays (constraining
different matrix elements, for the $b$ and $c$ quarks), and from
$\pi^0$ decays (giving access to a smaller phase space and, again,
for different matrix elements).

Using $58\times10^{6}~J/\psi$ events, the BESII experiment obtained
a first upper limit $\mathcal{B}(\eta (\eta^\prime)
\to\rm{invisible})/\mathcal{B}(\eta\, (\eta^\prime)
\to\gamma\gamma)<1.65\times10^{-3}~(6.69\times10^{-2})$,
corresponding to $\mathcal{B}(\eta\,
(\eta^\prime)\to\rm{invisible})<6.5\times10^{-4}
~(1.5\times10^{-3})$~\cite{np5}. Complementary to the BESII results,
IceCube set
$\mathcal{B}(\eta\to\nu_{e,\tau}\bar{\nu}_{e,\tau})<6.1\times10^{-4}$~\cite{icecube}
for $\eta$ decays into SM neutrinos. We present here updated results
of  searches for the invisible decays of $\eta$ and $\eta^\prime$.
The data sample used consists of $(225.3\pm2.8)\times10^{6}$
$J/\psi$ events \cite{njpsi} collected with the BESIII
detector~\cite{besnim} at the BEPCII collider~\cite{bepcii}.

%%%%%%%%%%%%%%%
%
% End of introduction part
%%%%%%%%%%%%%%%%%%%%%%%%%%%%%%%

\section{The BESIII Experiment and Monte Carlo simulation}
\label{sec:detector}

BEPCII/BESIII~\cite{besnim} is a major upgrade of the BESII
experiment at the BEPC accelerator.
%~\cite{besii}.
The design peak
luminosity of the double-ring $e^+e^-$ collider, \nohyphens{BEPCII}, is
$10^{33}$ cm$^{-2}$ s$^{-1}$ at a beam current of 0.93 A. The \nohyphens{BESIII}
detector has a geometrical acceptance of 93\% of $4\pi$ and consists
of four main components: (1) a small-celled, helium-based main draft
chamber (MDC) with 43 layers, which provides measurements of
ionization energy loss ($dE/dx$). The average single wire resolution is 135
$\mu$m, and the momentum resolution for charged particles with momenta of 1 GeV/$c$
in a 1 T magnetic field is 0.5\%; (2) an electromagnetic calorimeter
(EMC) made of 6240 CsI (Tl) crystals arranged in a cylindrical shape
(barrel) plus two end-caps. For 1.0 GeV photons, the energy
resolution is 2.5\% in the barrel and 5\% in the end-caps, and the
position resolution is 6 mm in the barrel and 9 \nohyphens{mm} in the end-caps;
(3) a time-of-flight system (TOF) for particle identification (PID)
composed of a barrel part made of two layers with 88 pieces of 5 cm
thick, 2.4 m long plastic scintillators in each layer, and two
end-caps with 96 fan-shaped, 5 cm thick, plastic scintillators in
each end-cap. The time resolution is 80 ps in the \nohyphens{barrel}, and 110 ps
in the end-caps, corresponding to a 2$\sigma$ K/$\pi$ separation for
momenta up to about 1.0 GeV/$c$; (4) a muon chamber system made of
1000 m$^2$ of resistive-plate-chambers arranged in 9 layers in the
barrel and 8 layers in the end-caps and incorporated in the return
iron of the super-conducting magnet. The position resolution is
about 2 cm.

The optimization of the event selection and the estimation of
physics backgrounds are performed using Monte Carlo (MC) simulated
data samples. The {\sc geant4}-based simulation software
BOOST~\cite{geant4} includes the geometric and material description
of the BESIII detectors, the detector response and digitization
models, as well as the tracking of the detector running conditions
and performance. The production of the $J/\psi$ resonance is
simulated by the MC event generator {\sc kkmc}~\cite{kkmc}; the
known decay modes are generated by {\sc evtgen}~\cite{evtgen} with
branching \nohyphens{ratios} set at PDG values~\cite{pdg}, while the remaining unknown
decay modes are modeled by {\sc lundcharm}~\cite{lund}.

\section{Data Analysis }
\label{sec:selection}

\subsection{\boldmath Analyses for $\eta$ and $\eta^\prime \to \text{invisible}$  }
\label{sec:invisible:selection}

In order to detect invisible $\eta$ and $\eta^\prime$ decays, we use
$J/\psi \to \phi \eta$ and $\phi\eta^\prime$. These two-body decays
provide a very simple event topology, in which the $\phi$
\nohyphens{candidates} can be reconstructed easily and cleanly
decaying into $K^+K^-$. The reconstructed $\phi$ particles can be used
to tag $\eta$ and $\eta^\prime$ in order to allow a search for their
invisible decays.  In addition, both the $\phi$ and
$\eta(\eta^\prime)$ are given strong boosts in the $J/\psi$ decay, so
the directions of the $\eta$ and $\eta^\prime$ decays are well defined
in the lab system and any decay products can be efficiently detected
by the BESIII detector. The missing $\eta$ and $\eta^\prime$ can be
searched for in the distribution of mass recoiling against the $\phi$
candidate.

Charged tracks in the BESIII detector are reconstructed using
track-induced signals in the MDC. We select tracks that originate
within $\pm10$ cm of the interaction point (IP) in the beam
direction and within 1 cm in the plane perpendicular to the beam.
The tracks must be within the MDC fiducial volume, $|\cos\theta| <
0.93$ ($\theta$ is the polar angle with respect to the $e^+$ beam
direction). Candidate events are required to have only \nohyphens{two} charged tracks
reconstructed with a net charge of \nohyphens{zero}. For each charged track,
information from TOF and $dE/dx$ are combined to calculate
$\chi^{2}_{\rm{PID}}(i)$  values. With the corresponding number of
degree of freedom, we obtain probabilities, $\rm{Prob}_{\rm{PID}}(i)$,
%and corresponding confidence levels $\rm{Prob}_{\rm{PID}}(i)$
for the hypotheses that a track is a pion, kaon, or proton, where
$i$ ($i=\pi/K/p$) is the particle type. For both kaon candidates, we
require $\rm{Prob}_{\rm{PID}}(K)> \rm{Prob}_{\rm{PID}}(\pi)$. The
mass recoiling against the $\phi$ candidate,
$M^{\text{recoil}}_{\phi}$, is calculated using the four-momentum of
the incident beams in the lab frame $(p^\mu_{\text{lab}} =
p^\mu_{e^-} + p^\mu_{e^+})$, and constructing the 4-product
$(M^{\text{recoil}}_{\phi})^2 = (p_{\text{lab}} -
p_{KK})^\mu(p_{\text{lab}} - p_{KK})_\mu$, where
$p^\mu_{KK}=p^\mu_{\phi}$ is the sum of the four-momentum of the two
charged kaons. The $\eta$ and $\eta^\prime$ signal regions in the
$M^{\text{recoil}}_{\phi}$ distribution are defined to be within
$3\sigma$ of the known masses of $\eta$ and
$\eta^\prime$~\cite{pdg}. Here, $\sigma$ is  the detector resolution
and is 17.8 (9.3) MeV/$c^2$, which is determined from MC simulation,
for $J/\psi\to\phi\eta (\eta^\prime)$.
\begin{figure*}[hbtp]
\begin{center}
\includegraphics[width=0.47\linewidth]{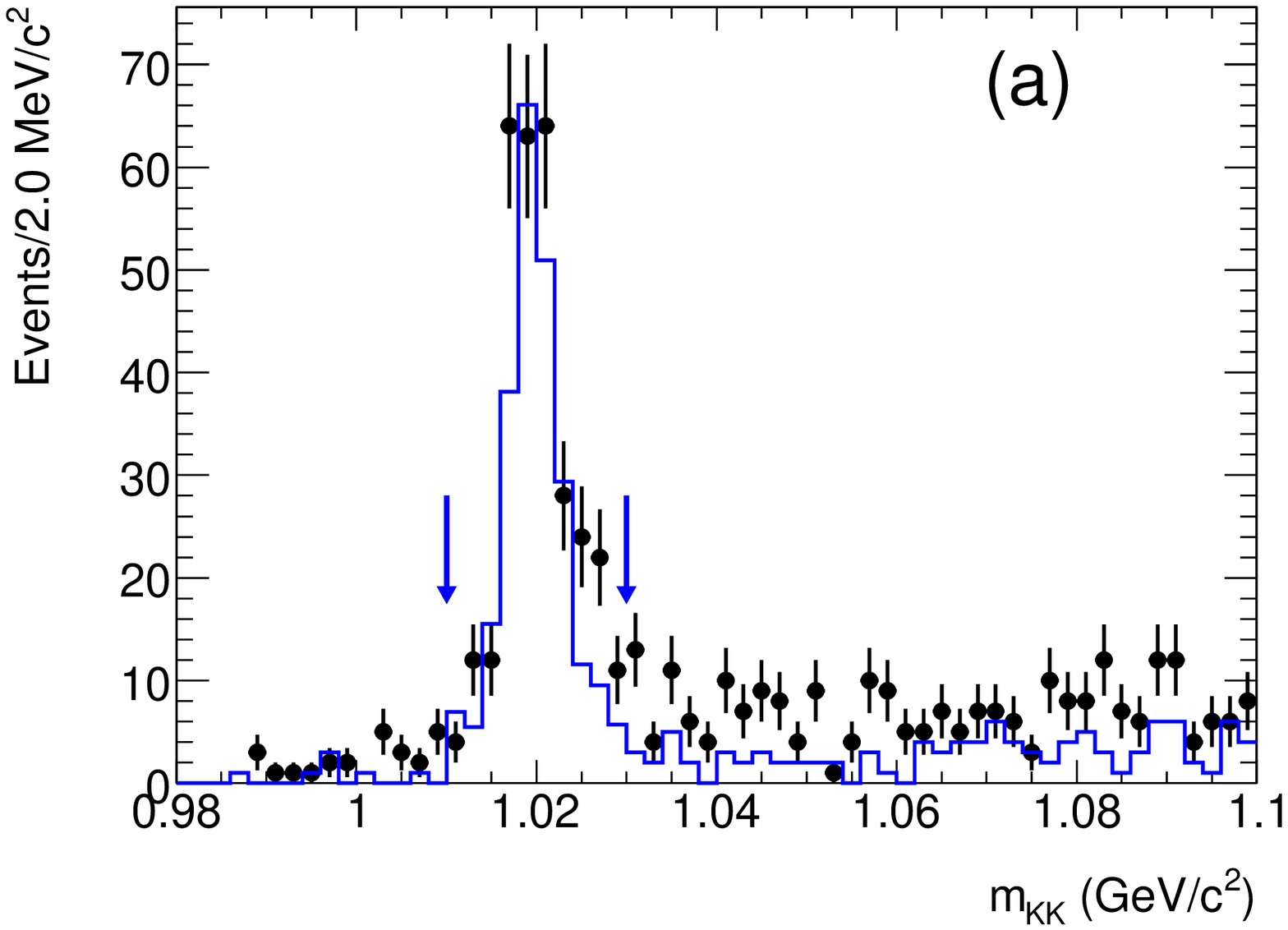}
  \includegraphics[width=0.47\linewidth]{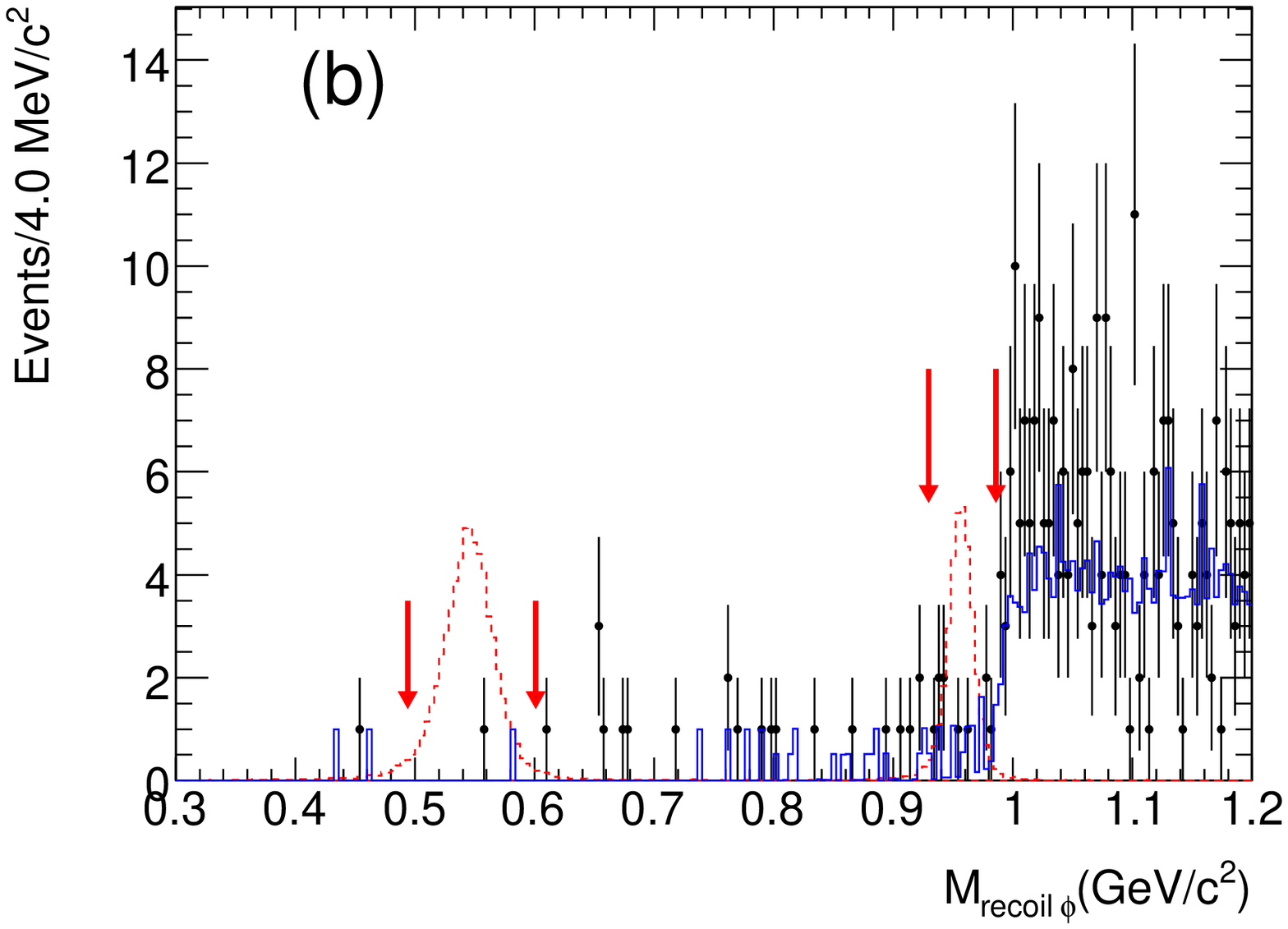}
  \caption{(a) The $m_{KK}$ distribution for candidate
events in data. The arrows on the plot indicate the signal region of
$\phi$ candidates. Points with error bars are
  data; the (blue) histogram is expected background. (b) Recoil mass distribution against $\phi$
candidates,
  $M^{\text{recoil}}_{\phi}$, for events with $1.01$ GeV/$c^2$ $< m_{KK} <1.03$ GeV/$c^2$ in (a). Points with error bars are
  data;
  the (blue) solid histogram is the sum of the expected backgrounds;
  the dashed histograms (with arbitrary scale) are signals of $\eta$ and
  $\eta^\prime$ invisible decays from MC simulations; the arrows on the plot
indicate the signal regions of the $\eta$ and $\eta^\prime \to
\text{invisible}$.}
  \label{eta-etap-invidata}
\end{center}
\end{figure*}

\begin{figure}[hbtp]
\begin{center}
  \includegraphics[width=0.825\linewidth]{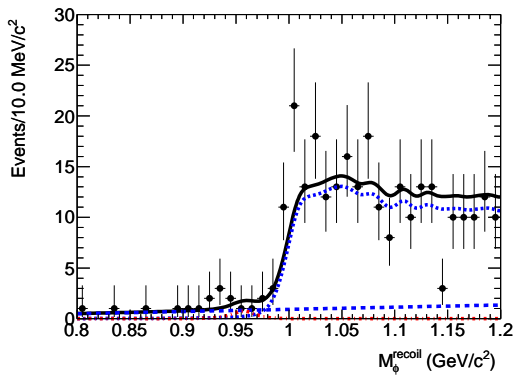}
  \caption{The $M^{\text{recoil}}_\phi$ distribution with events around the $\eta^\prime$ mass region.
Points with error bars are data.
 The (black) solid curve shows the result of the fit to
signal plus background distributions, the (blue) dotted curve shows
the background shape from $J/\psi\to\phi f_{0}(980)(f_{0}(980)\to
K_L K_L)$,  the (blue) dashed curve shows the polynomial function
for $J/\psi\to\phi K_L K_L$ background, and the (red) dotted-dash
curve shows the signal yield.}
  \label{etapinvidata}
\end{center}
\end{figure}
\begin{figure}[hbtp]
\begin{center}
  \includegraphics[width=0.825\linewidth]{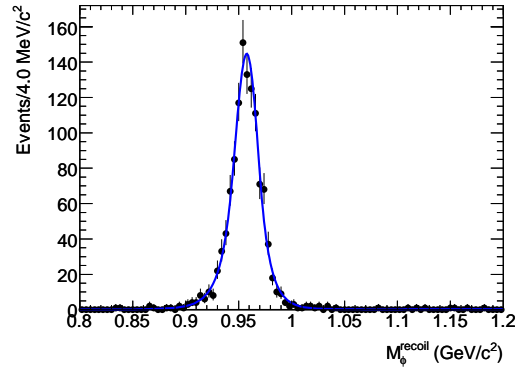}
  \caption{The $M^{\text{recoil}}_\phi$ distribution for the control sample
  $J/\psi \to \phi
\eta^\prime$, $\eta^\prime \to \pi^+\pi^- \eta (\eta\to \gamma
\gamma)$ decay candidates. The solid curve shows the fit results.}
  \label{fig:control}
\end{center}
\end{figure}

 Electromagnetic showers are
reconstructed from clusters of energy deposits in the EMC crystals.
The shower energies are required to be greater than $25$ MeV for the
barrel region ($|\cos\theta|<0.8$) and $50$ MeV for the end-cap
region ($0.86<|\cos\theta|<0.92$). The showers in the transition
region between barrel and end-cap are required to have an energy
greater than $100$ MeV. Showers must be isolated from all charged
tracks by more than $10^\circ$.

We require that $\eta(\eta^\prime) \to$ invisible events have no
charged tracks besides those of the $\phi \to K^+K^-$ candidate. In
addition, the number of EMC showers ($N_{\text{shower}}$), that
could be from a $K_L$ or a photon, are required to be zero inside a
cone of 1.0 rad around the recoil direction against the $\phi$
candidate. This requirement rejects most $\eta$ and $\eta^\prime$
decays into visible final states. It also eliminate most backgrounds
from multibody decays of $J/\psi \to \phi+$anything. In order to
ensure that $\eta$ and $\eta^\prime$ decay particles are inside the
fiducial volume of the detector, the recoil direction against the
$\phi$ is required to be within the region
$|\cos\theta_{\text{recoil}}|<0.7$, where $\theta_{\text{recoil}}$
is the polar angle of the recoil three-momentum of $\phi$ candidate.
Figure~\ref{eta-etap-invidata} (a) shows the $K^+K^-$ invariant mass
distribution after the above selection. A clear $\phi$ peak is seen.
Figure~\ref{eta-etap-invidata} (b) shows the recoil mass against
$\phi$ candidates for events with $1.01$ GeV/$c^2$ $< m_{KK} <1.03$
GeV/$c^2$, and there are no significant signals in the $\eta$ and
$\eta^\prime$ mass regions.

We use MC simulated events to determine selection \nohyphens{efficiencies} for the
signal channels and study possible backgrounds. The efficiencies are
$36.0\%$ and $36.1\%$ for $\eta$ and $\eta'$ invisible decays,
respectively.
% Stimulated by preliminary background studies of the
%$J/\psi$ inclusive MC sample,
More than 20 exclusive decay modes are generated with full
\nohyphens{MC} simulations in order to better understand the
backgrounds. The sources of backgrounds are divided into two
classes. Class I: The background is from $J/\psi \to \phi \eta
(\eta^\prime)$, where $\phi \to K^+K^-$ and $\eta(\eta^\prime)$
decays into visible final states that are not detected by the EMC.
The expected number of background events from this class is
$0.18\pm0.02~(1.0\pm0.2)$ in the signal region for the
$\eta$($\eta^\prime$) case. Class II: It is from $J/\psi$ decays to
final states without $\eta(\eta^\prime)$ or without both
$\eta(\eta^\prime)$ and $\phi$. For the $\eta$ invisible decay, the
dominant background is from $J/\psi \to \gamma\eta_{c},~\eta_c\to
K^{\pm}\pi^{\mp}K_L$, where the soft radiative photon is either
undetected or outside of the 1 rad cone against recoil $\phi$
direction in the EMC and the fast $\pi$ is mis-identified as kaon.
We determine the expected number of background from $J/\psi \to
\gamma\eta_{c},~\eta_c\to K^{\pm}\pi^{\mp}K_L$ with a phase space
distribution for the $\eta_c\to K^{\pm}\pi^{\mp}K_L$ decay in MC
simulation, and a systematic uncertainty is assigned to cover the
variation due to possible structures on the Dalitz plot.  For the
$\eta^\prime$ case, the dominant background is from $J/\psi \to \phi
K_L K_L$ and $J/\psi \to \phi f_{0}(980),~f_{0}(980)\to K_L K_L$.
The expected number of background events from class II is $0.8 \pm
0.2$ and $9.4\pm 1.7$ in the signal regions for $\eta$ and
$\eta^\prime$, respectively.

After all selection criteria are applied, only one \nohyphens{event}
(shown in Fig.~\ref{eta-etap-invidata} (b)) survives in the $\eta$
signal region where $1.0\pm 0.2$ background event is expected. An
upper limit (UL) at the $90\%$ confidence level (C.L.) of
$N^\eta_{\text{UL}}=3.34$ for $J/\psi \to \phi \eta$ ($\phi \to
K^+K^-$ and $\eta \to \text{invisible}$) is obtained using the
POLE$^{++}$ program~\cite{pole} with the Feldman-Cousins frequentist
approach \cite{feldmancousin}. The information used to obtain the
upper limit includes the number of observed events in the signal
region, and the expected number of background events and their
uncertainty.

For the $\eta^\prime$ case, an unbinned extended maximum likelihood
(ML) fit to the $M^{\text{recoil}}_\phi$ distribution in the range
$0.8$ GeV/$c^2$ $<M^{\text{recoil}}_\phi<1.2$ GeV/$c^2$, as shown in
Fig.~\ref{etapinvidata}, is performed. The signal shape used in the
fit, shown in Fig.~\ref{fig:control}, is obtained from a nearly
background-free $J/\psi \to \phi \eta^\prime$, $\eta^\prime \to
\pi^+\pi^- \eta,~\eta\to \gamma \gamma$ sample. The \nohyphens{purity} of the
sample is greater than 98.5\%. The shape of the invisible signal peak
in the $M^{\text{recoil}}_\phi$ distribution is fixed to the smoothed
histograms of the $J/\psi \to \phi \eta^\prime$, $\eta^\prime \to
\pi^+\pi^- \eta,~\eta\to \gamma \gamma$ MC sample, and the signal yield
is allowed to float.  The shape of the dominant background $J/\psi \to
\phi f_0(980)$, $f_0(980) \to K_L K_L$ is described by MC simulated
data, in which the $f_0(980)$ line shape is parameterized with the
Flatt$\acute{\text{e}}$ form~\cite{flatte}
\begin{eqnarray}
f(m) = \frac{1}{M^2_{f_0} + m^2 + i  (g^2_1\rho_{\pi\pi} +
g^2_2\rho_{KK})}, \label{eq:flatte}
\end{eqnarray}
where $M_{f_0}$ is the mass of the $f_0(980)$, $m$ is the effective
mass, $\rho$ is Lorentz invariant phase space ($\rho=2k/m$, here,
$k$ refers to the $\pi$ or $K$ momentum in the rest frame of the
resonance), and $g_1$ and $g_2$ are coupling-constants for the
$f_0(980)$ resonance coupling to the $\pi\pi$ and $KK$ channels,
respectively.
 These parameters
[$M_{f_0}= 0.965\pm 0.010$ GeV/$c^2$, $g^{2}_1= 0.165\pm0.018$
(GeV/$c^2$)$^2$ and $g^{2}_2= 0.695\pm 0.075$ (GeV/$c^2$)$^2$] have
been determined in the analysis of $J/\psi \to \phi \pi^+\pi^-$ and
$\phi K^+K^-$ from BESII data~\cite{pwabes2,pwabes2-1}. In the ML
fit, the dominant background shape ($J/\psi \to \phi f_0(980)$,
$f_0(980) \to K_L K_L$) is fixed to the MC simulations, and its yield
($N^{\text{bkg}}_{f_0}$) is floated. The shape of the remaining
background from $J/\psi \to \phi K_L K_L$ is modeled with a first
order Chebychev polynomial whose slope and yield
($N^{\text{bkg}}_{\text{non-}f_0}$) are floated in the fit to data.
 The signal yield,
$N^{\eta^\prime}_{\text{sig}} = 2.3\pm 4.3$, is consistent with zero
observed events, and the resulting fitted values of
$N^{\text{bkg}}_{f_0}$ and $N^{\text{bkg}}_{\text{non-}f_0}$ are
$239\pm28$ and $37\pm25$, respectively, where the errors are
statistical. We obtain an upper limit by \nohyphens{integrating} the normalized
likelihood distribution over the positive values of the number of
signal events. The upper limit at the 90\% C.L. is
$N^{\eta^\prime}_{UL} = 10.1$.

\subsection{\boldmath Analyses for $\eta$ and $\eta^\prime \to \gamma \gamma $  }
\label{sec:gg:selection}

The branching fraction of $\eta(\eta^\prime) \to \gamma \gamma$ is
also determined in $J/\psi \to \phi \eta (\eta^\prime)$, in
order to obtain the ratio of $\mathcal{B}(\eta(\eta^\prime)\to
\text{invisible})$ to $\mathcal{B}(\eta(\eta^\prime)\to \gamma
\gamma)$. The advantage of measuring
$\frac{\mathcal{B}(\eta(\eta^\prime)\to
\text{invisible})}{\mathcal{B}(\eta(\eta^\prime)\to \gamma \gamma)}$
is that the uncertainties due to the total number of $J/\psi$
events, tracking efficiency, PID, the number of the charged tracks,
and the residual noise in the EMC cancel.

The selection criteria for the charged tracks are the same as those
for $J/\psi \to \phi \eta (\eta^\prime)$, $\eta(\eta^\prime)\to
\text{invisible}$. However, at least two good photons are
required. The events are kinematically fitted using energy and
momentum conservation constraints (4C) under the $J/\psi \to
KK\gamma\gamma$  \nohyphens{hypothesis} in order to obtain better mass
resolution and suppress backgrounds further. We require the
kinematic fit $\chi^{2}_{K^{+}K^{-}\gamma\gamma}$ to be less than 90 (40)
for the $\eta(\eta^\prime)$ case. If there are more than two
photons, the fit is repeated using all permutations, and the
combination with the best fit to $KK\gamma\gamma$ is retained.

\begin{figure}[hbtp]
\begin{center}
  \includegraphics[width=0.8\linewidth]{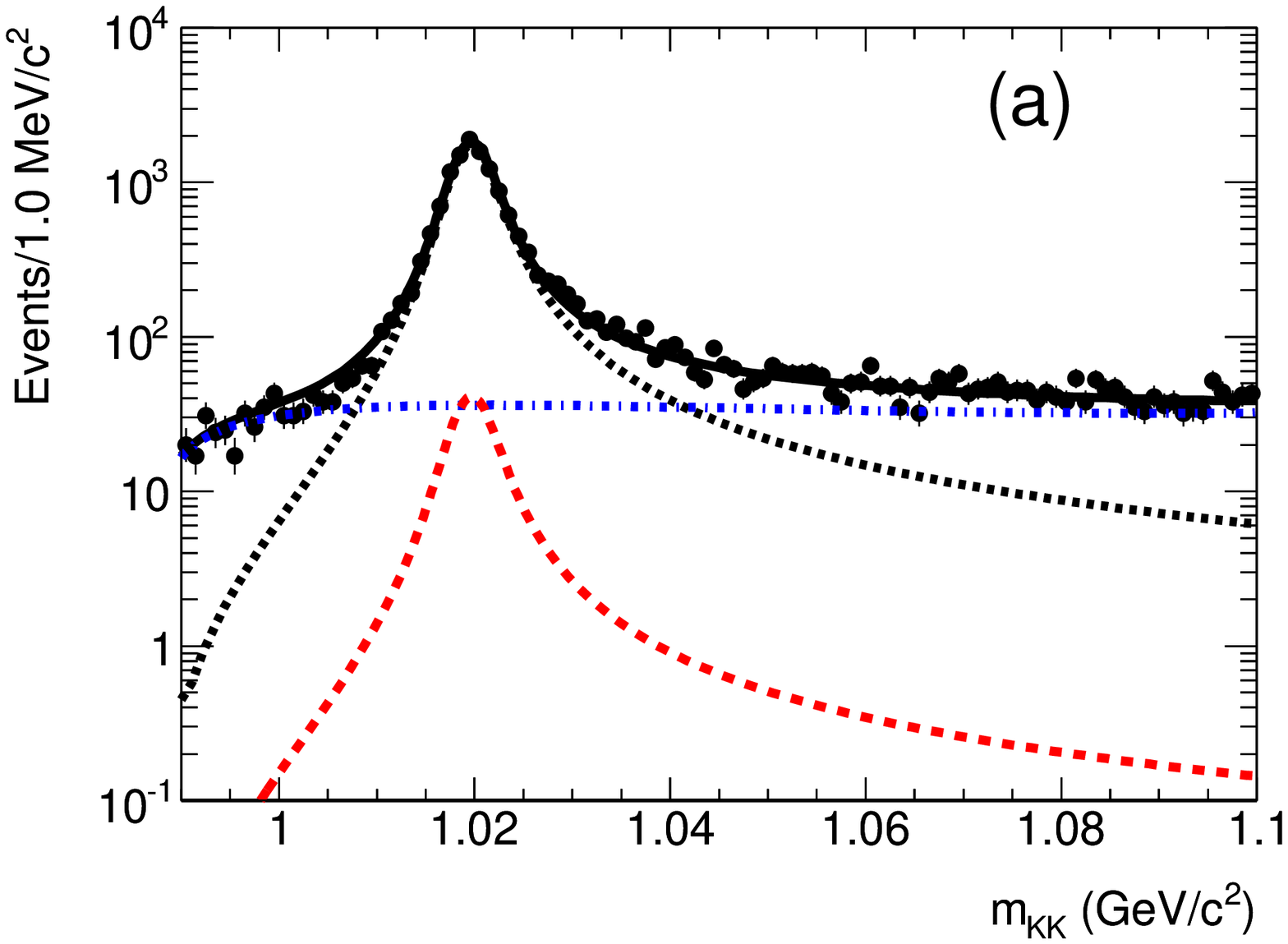}
  \includegraphics[width=0.8\linewidth]{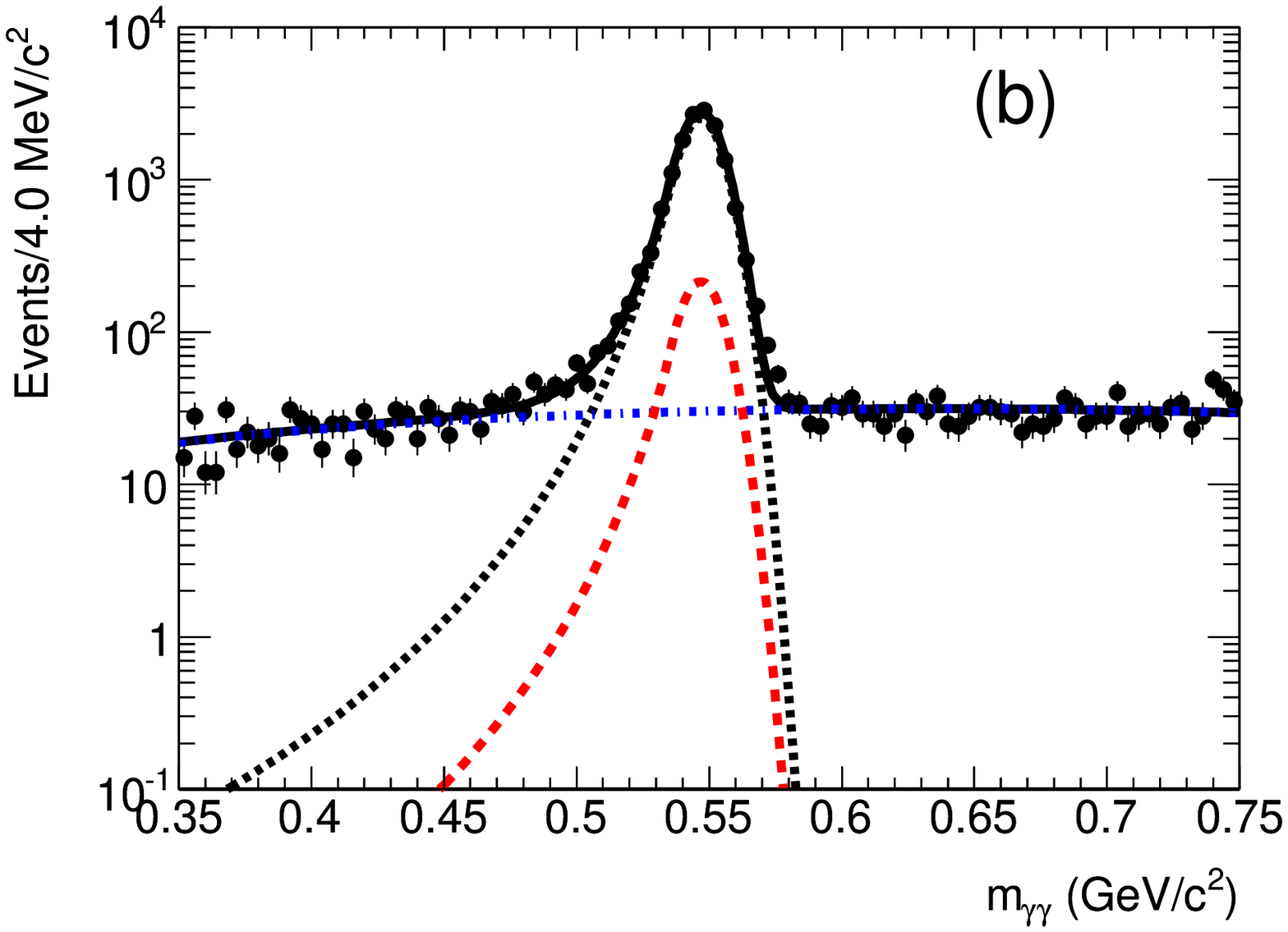}
  \caption{The (a) $m_{KK}$ and (b) $m_{\gamma\gamma}$
  distributions with fit results superimposed for $J/\psi \to \phi \eta$, $\phi \to K^+K^-$, $\eta\to \gamma\gamma$. Points with error bars are
  data. The (black) solid curves show the results of the fits to
signal plus background, and the (black) dashed curves are
for signal.  In (a), the (blue) dotted-dash curve shows
non-$\phi$-peaking backgrounds, and the (red) short-dashed curve shows the
non-$\eta$-peaking background. In (b), the (blue) dotted-dash curve
shows non-$\eta$-peaking backgrounds, and the (red) short-dashed curve
shows the non-$\phi$-peaking  background.}
  \label{figetafit}
\end{center}
\end{figure}

\begin{figure}[hbtp]
\begin{center}
  \includegraphics[width=0.8\linewidth]{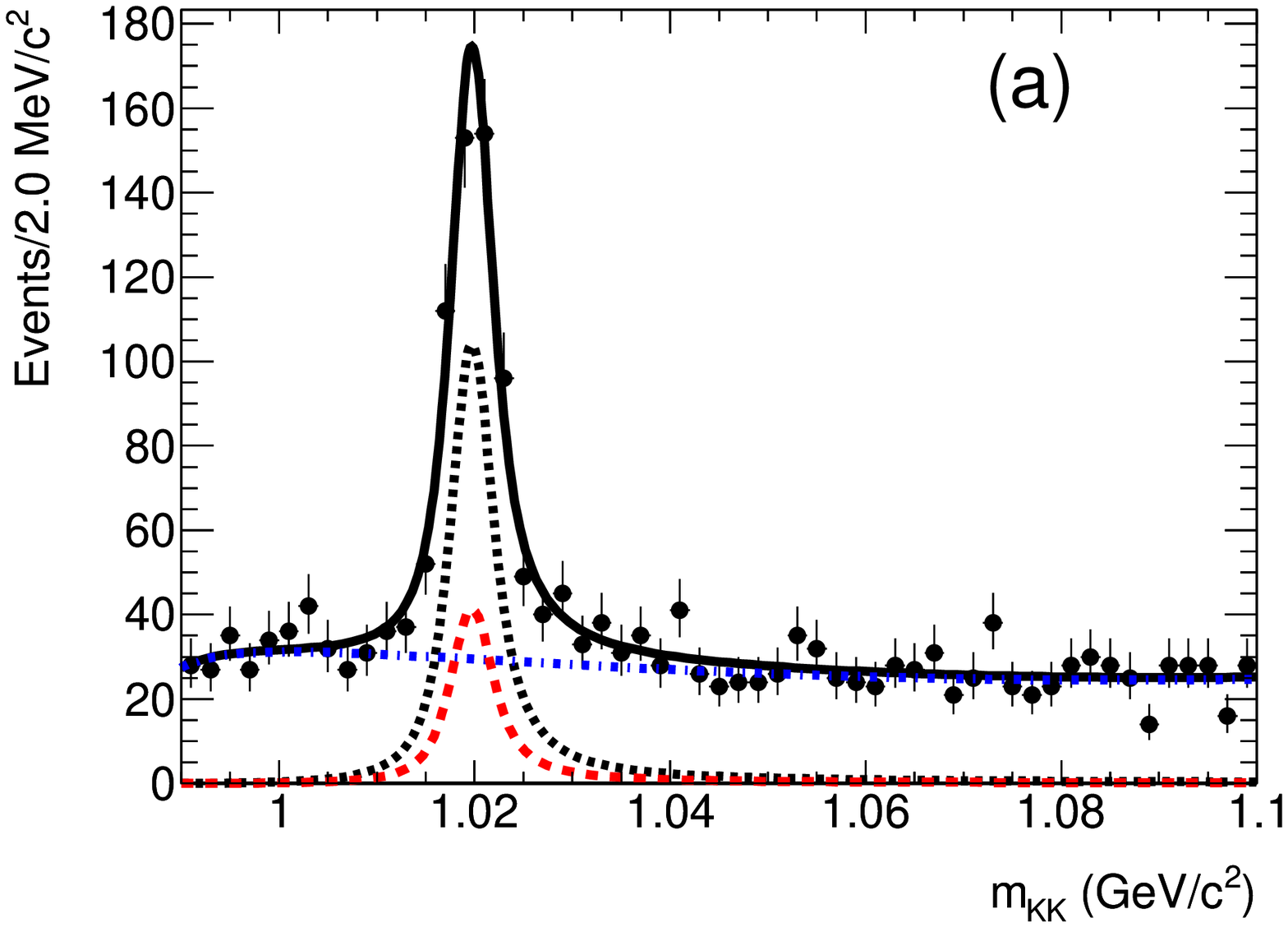}
  \includegraphics[width=0.8\linewidth]{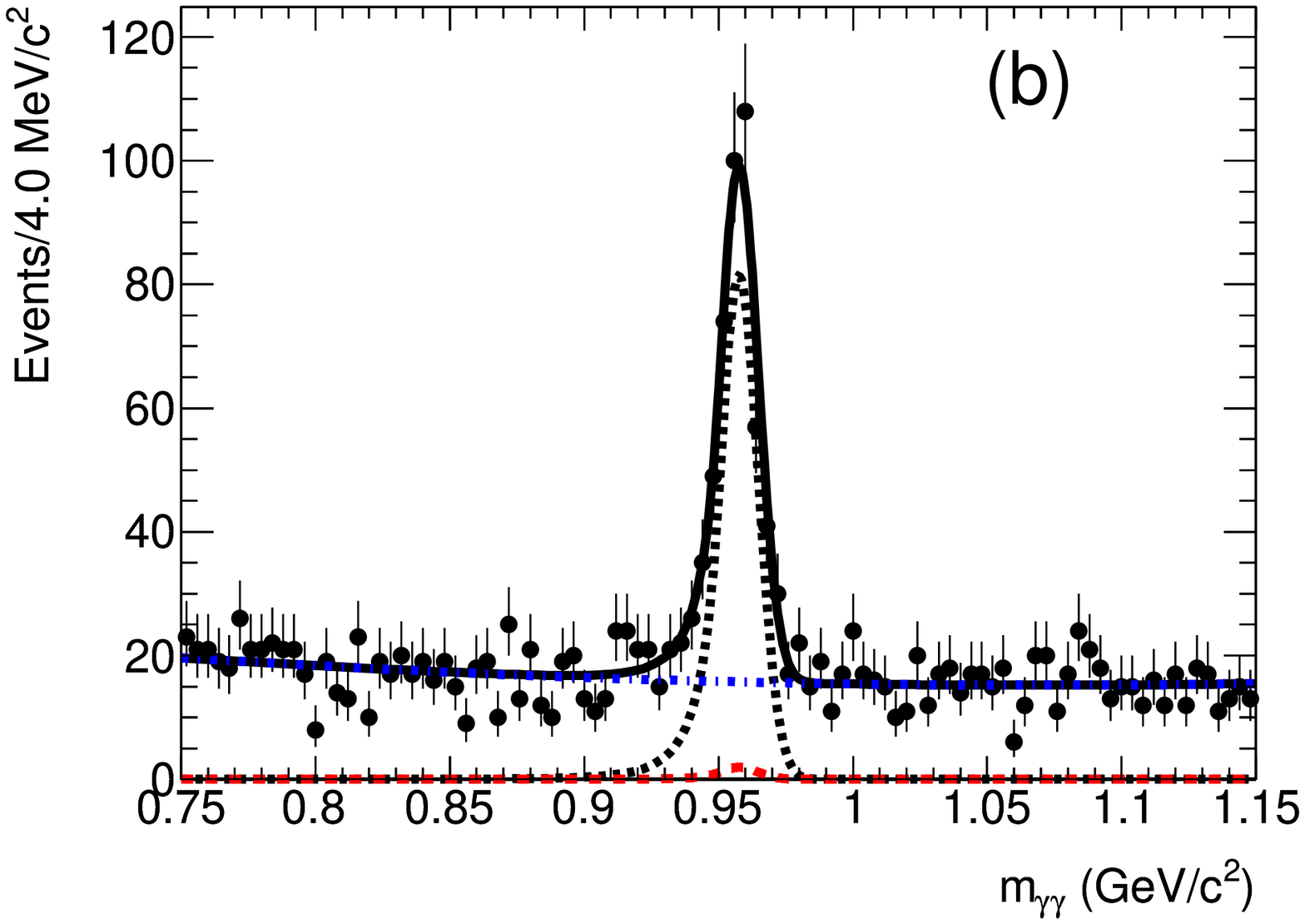}
  \caption{The (a) $m_{KK}$ and (b) $m_{\gamma\gamma}$
  distributions with fit results superimposed for
  $J/\psi \to \phi \eta^\prime$, $\phi \to K^+K^-$, $\eta^\prime\to \gamma\gamma$.
  Points with error bars are
  data. The (black) solid curves show the results of the fits to
signal plus background distributions, and the (black) dashed curves are
for signal. In (a), the (blue) dotted-dash curve shows
non-$\phi$-peaking backgrounds, and the (red) short-dashed curve shows the
non-$\eta^\prime$-peaking background. In (b), the (blue) dotted-dash
curve shows non-$\eta^\prime$-peaking backgrounds, and the (red)
short-dashed curve shows the non-$\phi$-peaking  background.}
  \label{figetapfit}
\end{center}
\end{figure}

The numbers of $J/\psi\to\phi\eta(\eta^\prime),$
$\eta(\eta^\prime)\to\gamma\gamma$ events are obtained from an
extended unbinned ML fit to the $K^{+}K^{-}$ {\it versus}
$\gamma\gamma$ invariant mass distributions. The projection of the fit
on the $m_{KK}$ ($m_{\gamma\gamma}$) axis is shown in
Figs.~\ref{figetafit}(a) and~\ref{figetapfit}(a)
(Figs.~\ref{figetafit}(b) and ~\ref{figetapfit}(b)) for the $\eta$ and
$\eta^\prime$ cases, respectively. In the ML fits, we require that
$0.99$ GeV/$c^2$ $<m_{KK} <1.10$ GeV/$c^2$ and $0.35$ GeV/$c^2$ $<
m_{\gamma \gamma}<0.75$ GeV/$c^2$ ($0.75$ GeV/$c^2$ $< m_{\gamma
  \gamma}<1.15$ GeV/$c^2$) for the $\eta(\eta^\prime)$ case. The signal
shape for $\phi$ is modeled with a relativistic Breit-Wigner ($RBW$)
function~\cite{rbw} convoluted with a Gaussian function that
represents the detector resolution; the signal shape for
$\eta(\eta^\prime)$ is described by a Crystal Ball ($CB$)
function~\cite{cbf}, and its parameters are floated. In the ML fits,
the width of $\phi$ is fixed at the PDG value, and its central mass
value is floated. The backgrounds are divided into three categories:
non-$\phi\eta(\eta^\prime)$-peaking background ({\it i.e.,} $J/\psi
\to \gamma \pi^0 K^+K^-$, in which one of the photons is missing);
non-$\phi$-peaking background ({\it i.e.,} $J/\psi \to K^+K^- \eta
(\eta^\prime)$); and non-$\eta(\eta^\prime)$-peaking background ({\it
  i.e.,} $J/\psi \to \phi \gamma \gamma$ and $\phi \pi^0\pi^0$ ). The
probability density functions (PDF) for non-$\phi$-peaking background
in the $m_{KK}$ distribution is parameterized by~\cite{kkbgfunction}
\begin{eqnarray}
B(m_{KK}) = (m_{KK}-2m_{K})^{a} \cdot e^{-bm_{KK}-cm^{2}_{KK}},
\label{bgphipdf}
\end{eqnarray}
where $a$, $b$ and $c$ are free parameters, and $m_K$ is the nominal
mass value of the charged kaon from the PDG~\cite{pdg}. The shape for
the non-$\eta(\eta^\prime)$-peaking background in the $m_{\gamma
\gamma}$ distribution is modeled by a second-order Chebychev
polynomial function ($B(m_{\gamma\gamma})$). All parameters related
to the background shape are floated in the fit to data.   The PDFs
for signal and backgrounds are combined in the likelihood function
$\mathcal{L}$, defined as a function of the free parameters
$N^\eta_{\gamma\gamma}$, $N^{\text{non-}\phi\eta}_{\text{bkg}}$,
$N^{\text{non-}\phi}_{\text{bkg}}$, and
$N^{\text{non-}\eta}_{\text{bkg}}$:
\begin{eqnarray}
\mathcal{L} &=&
\frac{e^{-(N^\eta_{\gamma\gamma}+N^{\text{non-}\phi\eta}_{\text{bkg}}+N^{\text{non-}\phi}_{\text{bkg}}+N^{\text{non-}\eta}_{\text{bkg}})}}
{N!} \nonumber \\
&&\times \prod^N_{i=1}[N^\eta_{\gamma\gamma}RBW(m^i_{KK})\times
CB(m^i_{\gamma\gamma}) \nonumber\\
&&+ N^{\text{non-}\phi\eta}_{\text{bkg}} B(m^i_{KK})\times
B(m^i_{\gamma\gamma}) \nonumber\\
&&+ N^{\text{non-}\phi}_{\text{bkg}} B(m^i_{KK})\times CB(m^i_{\gamma\gamma}) \nonumber\\
&&+ N^{\text{non-}\eta}_{\text{bkg}} RBW(m^i_{KK})\times
B(m^i_{\gamma\gamma})], \label{eq:pdf:2d:gg}
\end{eqnarray}
where $N^\eta_{\gamma\gamma}$ is the number of $J/\psi \to \phi
\eta,~\phi \to K^+K^-,~\eta \to \gamma \gamma$ events, and
$N^{\text{non-}\phi\eta}_{\text{bkg}}$,
$N^{\text{non-}\phi}_{\text{bkg}}$, and
$N^{\text{non-}\eta}_{\text{bkg}}$ are the numbers of the
corresponding three kinds of backgrounds. The fixed parameter $N$ is
the total number of selected events in the fit region, and
$m^i_{KK}$ ($m^i_{\gamma\gamma}$) is the value of $m_{KK}$
($m_{\gamma\gamma}$) for the $i$th event. We use the product of the
PDFs, since we have verified that $m_{KK}$ and $m_{\gamma\gamma}$
are uncorrelated for each component. The negative log-likelihood
($-\text{ln}\mathcal{L}$) is then minimized with respect to the
extracted yields.   The resulting fitted signal and background
yields are summarized in Table~\ref{tab:ggresults}. We also obtain
the results for the $\eta^\prime$ case by replacing $\eta$ with
$\eta^\prime$ in Eq.~(\ref{eq:pdf:2d:gg}).
  The fitted results for $\eta (\eta^\prime)
\to \gamma\gamma$ are shown in Fig.~\ref{figetafit}
(Fig.~\ref{figetapfit}). The detection efficiencies are determined
with MC simulations to be $36.3\%$ and $31.7\%$ for $\eta$ and
$\eta^\prime$, respectively.
\begin {table}[htbp]
\begin {center}\small
\caption{The fitted signal and background yields for
$J/\psi\to\phi\eta(\eta^\prime)$,
$\eta(\eta^\prime)\to\gamma\gamma$, and
$\epsilon^\eta_{\gamma\gamma}(\epsilon^{\eta^\prime}_{\gamma\gamma})$
is its selection efficiency.  } \label{tab:ggresults}
\renewcommand{\arraystretch}{1.4}
\begin {tabular}{l|cc}
\hline \hline
 & \multicolumn{2}{c}{Value} \\
 Quantity            & $\eta$                 & $\eta^\prime$  \\
\hline
$N^\eta_{\gamma\gamma}(N^{\eta^\prime}_{\gamma\gamma})$            &    $13390 \pm 136$       &  $400 \pm 25$    \\
\hline
$N^{\text{non-}\phi\eta}_{\text{bkg}}(N^{\text{non-}\phi\eta^\prime}_{\text{bkg}})$        &        $2514\pm64$          &  $1482\pm46$ \\
\hline
$N^{\text{non-}\phi}_{\text{bkg}}(N^{\text{non-}\phi}_{\text{bkg}})$                          &     $1132 \pm70$                  &  $10\pm15$  \\
\hline
$N^{\text{non-}\eta}_{\text{bkg}}(N^{\text{non-}\eta^\prime}_{\text{bkg}})$                &   $313\pm 54$             &  $159\pm 26$  \\
\hline
$\epsilon^\eta_{\gamma\gamma}(\epsilon^{\eta^\prime}_{\gamma\gamma})$                    &        36.3\%             &   31.7\%       \\
\hline\hline
\end {tabular}
\end {center}
\end {table}

%According to the results in Table~\ref{tab:ggresults}, the ratio of
%$\mathcal{B}(J/\psi \to \phi \eta)$ to $\mathcal{B}(J/\psi \to \phi
%\eta^\prime)$ is determined to be $1.65\pm 0.12$, where the error is
%only statistical. Using the PDG values~\cite{pdg}, we obtain the
%ratio $\frac{\mathcal{B}(J/\psi \to \phi \eta)}{\mathcal{B}(J/\psi
%\to \phi \eta^\prime)}= 1.88\pm 0.38$, which is consistent with our
%result within error.
According to the results in Table~\ref{tab:ggresults}, the ratio of
$\mathcal{B}(J/\psi \to \phi \eta)$ to $\mathcal{B}(J/\psi \to \phi\eta^\prime)$,
is found to be \nohyphens{consistent} with the known value~\cite{pdg}.
The individual branching fraction is larger by 1.3(1.6)$\sigma$ with respect to
the average value listed in Ref.~\cite{pdg} for $\mathcal{B}(J/\psi \to \phi \eta(\eta^\prime))$, while it is consistent with Ref.~\cite{np5}.

\section{Systematic uncertainties }
\label{sec:sys:selection}

The contributions to the systematic error on the calculation of the
ratios are summarized in Table \ref{syserr}. The uncertainty, due to
the requirement of no neutral showers in the EMC inside the $1.0$
rad cones around the recoil direction against the $\phi$ candidate,
is estimated using the control sample of fully reconstructed
$J/\psi\to\phi\eta$, $\eta\to\gamma\gamma$ events. The ratios of
events with no extra photons to events without this requirement are
obtained for both data and MC simulation. The difference 0.3\% is
taken as the systematic error for both the $\eta$ and $\eta^\prime$
cases.  This study determines the difference of the noise in the EMC
for MC simulation and data. The uncertainty due to the $\phi$ mass
window requirement is determined to be 1.5\% by using the same
control sample of $J/\psi\to\phi\eta$, $\eta\to\gamma\gamma$ events.

For the $\eta$ invisible decay,  the dominant background is from
$J/\psi \to \gamma\eta_{c},~\eta_c\to K^{\pm}\pi^{\mp}K_L$. The
expected number of the background is estimated with the MC
simulations using a phase space distribution for $\eta_c\to
K^{\pm}\pi^{\mp}K_L$.  The uncertainty to the background estimate
 that covers the variation of the Dalitz plot structures is
 studied using the data sample of $\psi^\prime \rightarrow \gamma
 \eta_c$, $\eta_c \rightarrow K_s K^\pm \pi^\mp$ events, which were
 from BESIII in Ref.~\cite{etaclineshape}.   The experimental data suggest that the $\eta_c \rightarrow K_s K^\pm \pi^\mp$
 decays predominantly via the scalar $K^*_0(1430)$ meson,  i.e., $\eta_c \rightarrow K^*_0(1430)
 \bar{K}$, which is consistent with the results from BABAR and Belle
 experiements~\cite{babar-eta-c,belle-eta-c}.
 After correction for detection efficiency,  the experimental Dalitz plot
 distribution in the $\eta_c \rightarrow K_s K^\pm \pi^\mp$ is used to reweight the
$\eta_c\to K^{\pm}\pi^{\mp}K_L$ simulation. The reweighting
increases the expected number of background events by 5\%, which
leads to a relative error of 1.2\% on $\eta \rightarrow$ invisible
decay.

For the $\eta^\prime$ invisible decay, systematic errors in the \nohyphens{ML}
fit originate from the limited number of events in the data sample
and from uncertainties in the  PDF parameterizations. Since the
signal shape is obtained from the  $J/\psi \to \phi \eta^\prime$,
$\eta^\prime \to \pi^+\pi^- \eta,~\eta \to \gamma\gamma$ events in the
data, the uncertainty due to the signal shape is negligible. The
uncertainty due to the background shape is estimated by
varying the PDF shape of the background in the ML fit. The shape of
the dominant background $J/\psi \to \phi f_0(980)$, $f_0(980) \to
K_L K_L$ is parameterized with the Flatt$\acute{\text{e}}$ form in
Eq.~(\ref{eq:flatte}).
%and $M_{f_0}$, $g_1$ and $g_2$ parameters are
%fixed to the measured values ($M_{f_0}= 965\pm 10$ MeV/$c^2$, $g_1=
%165\pm18$ MeV/$c^2$ and $g_2= 695\pm 75$ MeV/$c^2$) from
%Ref.~\cite{pwabes2} in the ML fit.
To estimate the uncertainty, we
change the central values of the parameters used in the fit by one standard
deviation of the measured values~\cite{pwabes2},  and find that the
relative error on $\eta^\prime\to \text{invisible}$ decay is 1.0\%.
The systematic uncertainty due to the choice of parameterization for
the shape of the background from $J/\psi \to \phi K_L K_L$ is
estimated by varying  the order of the polynomial in the fit; we find a
relative change on the invisible signal yield of 2.9\%, which is
taken as the uncertainty due to the background model.

The uncertainty in the determination of the number of observed
$J/\psi \to \phi \eta(\eta^\prime)$, $\phi \to K^+K^-$,
$\eta(\eta^\prime)\to \gamma \gamma$ events is also estimated. The
systematic error due to photon detection is determined to be $1\%$
for each photon \cite{detectgamma}. The uncertainty due to the 4C fit is
estimated to be 0.4\%(0.8\%) for the $\eta(\eta^\prime)$ case using the
control sample $J/\psi\to\pi^{0}K^{+}K^{-}$. In the fit to the $\phi$
mass distribution, the mass resolution is fixed to the MC
simulation; the level of possible discrepancy is determined with a
smearing Gaussian, for which a non-zero $\sigma$ would represent a
MC-data difference in the mass resolution. The uncertainty
associated with a difference determined in this way is 0.1\% (1.0\%)
for the $\eta(\eta^\prime)$ case. The systematic uncertainty due to the
choice of parameterization for the shape of the
non-$\phi\eta(\eta^\prime)$-peaking background is estimated by
varying the order of the polynomial in the fit; we find the relative
changes on the $\eta(\eta^\prime)$ signal yield of 0.1\%~(0.6\%),
which is taken as the uncertainty due to the background shapes. The
total systematic errors $\sigma^{\text{sys}}_{\eta}$ and
$\sigma^{\text{sys}}_{\eta^\prime}$ on the ratio are $2.8\%$ and
$4.1\%$ for $\eta$ and $\eta^\prime$, as summarized in Table
\ref{syserr}.
\begin {table}[hbtp]
\begin {center}\small
\renewcommand{\arraystretch}{1.4}
\caption{Summary of errors. The first five lines are relative
systematic errors for $J/\psi\to\phi\eta(\eta^\prime)$,
$\eta(\eta^\prime)\to\text{invisible}$. The next four lines are
relative systematic errors for $J/\psi\to\phi\eta(\eta^\prime)$,
$\eta(\eta^\prime)\to\gamma\gamma$. The second line from the bottom
is the relative statistical error of
$N^\eta_{\gamma\gamma}(N^{\eta^\prime}_{\gamma\gamma})$. }
\label{syserr}
\begin {tabular}{lcc}
\hline \hline
                                   & \multicolumn{2}{c}{Sys. error
                                   (\%)} \\
Source of uncertainties                 & $\eta$    &   $\eta^\prime$  \\
\hline
 Requirement on $N_{\text{shower}}$             &  0.3           &     0.3           \\
$\phi$ mass window              &  1.5          &     1.5           \\
$J/\psi \rightarrow \gamma \eta_c$, $\eta_c \rightarrow K_L
K^\pm\pi^\mp$ background & 1.2 & -
\\
Background shape of $J/\psi\to \phi f_0(980)$    & -    & 1.0  \\
Background shape of $J/\psi\to \phi K_L K_L$    & -    & 2.9  \\ \hline
4C fit for $\eta(\eta^\prime)\to \gamma\gamma$       &   0.4    &  0.8    \\
 Photon detection      &    2.0    &  2.0 \\
 Signal shapes for  $\eta(\eta^\prime)\to \gamma\gamma$            &  0.1        &  1.0    \\
Background shape for $\eta(\eta^\prime)\to \gamma\gamma$  &   0.1    &  0.6   \\
\hline
Total systematic errors            &      2.8              &    4.1         \\
\hline
Statistical error of $N^\eta_{\gamma\gamma}(N^{\eta^\prime}_{\gamma\gamma})$   &   1.0  &  6.0\\
\hline
Total errors              &  3.0   &  7.4  \\
\hline\hline
\end {tabular}
\end {center}
\end {table}

\section{Results}

The upper limit at the $90\%$ confidence level on the ratio of $\mathcal{B}(\eta \rightarrow
\text{invisible})$ to $\mathcal{B}(\eta \to \gamma \gamma)$ is
calculated with
\begin{equation}\label{eqratio}
 \frac{\mathcal{B}(\eta\to \text{invisible})}{\mathcal{B}(\eta\to\gamma\gamma)} \,<\,
 \frac{N^\eta_{UL}/\epsilon_{\eta} }{N^\eta_{\gamma\gamma}/ \epsilon^\eta_{\gamma\gamma}}
 \,  \frac{1}{1-\sigma_{\eta}} \, ,
\end{equation}
where $N^\eta_{UL}$ is the 90\% upper limit of the number of
observed events for $J/\psi \to \phi \eta$, $\phi\to K^+K^-$,
$\eta\to \text{invisible}$ decay, $\epsilon_\eta$ is the MC
determined efficiency for the signal channel,
$N^\eta_{\gamma\gamma}$
 is the number of events for the $J/\psi \to \phi \eta$, $\phi \to K^+K^-$, $\eta\to \gamma\gamma$,
 $\epsilon^\eta_{\gamma\gamma}$ is the MC determined \nohyphens{efficiency}, and
  $\sigma_\eta$ is
  the total error for the $\eta$ case from Table~\ref{syserr}.
  The upper limit on the ratio of $\mathcal{B}(\eta'\to\text{invisible}$) to $\mathcal{B}(\eta'\to\gamma\gamma)$ is obtained similarly.
  Since only the statistical
error is considered when we obtain the 90\% upper limit of the
number of events, to be conservative, $N^\eta_{UL}$ and
$N^{\eta^\prime}_{UL}$  are shifted up by one sigma of the
additional uncertainties ($\sigma_\eta$ or $\sigma_{\eta^\prime}$ ).

Thus, the upper limit of $2.6\times 10^{-4}~(2.4\times 10^{-2})$ on the ratio of
$\mathcal{B}(\eta(\eta^\prime) \to \text{invisible})$ and
$\mathcal{B}(\eta(\eta^\prime) \to \gamma\gamma)$ is obtained at the
90\% confidence level.

%%%%%%%%%%%%%%%%%%
%%
%%  Start of conclusion
%%%%%%%%%%%%%%%%%%%%%
\section{Conclusion}

In summary, the invisible decays of $\eta$ and $\eta^\prime$ are
searched for in the two-body decays $J/\psi\to\phi\eta$ and
$\phi\eta^\prime$ using $(225.3\pm2.8)\times10^{6}$ $J/\psi$ decays
collected with the BESIII detector.  We find no signal above
background for the invisible decays of $\eta$ and $\eta^\prime$ and
obtain upper limits at the $90\%$ C.L. of $2.6\times10^{-4}$ and
$2.4\times10^{-2}$ for $\frac{\mathcal{B}(\eta \to
  \text{invisible})}{\mathcal{B}(\eta \to\gamma\gamma)}$ and
$\frac{\mathcal{B}(\eta^\prime \to
  \text{invisible})}{\mathcal{B}(\eta^\prime \to\gamma\gamma)}$,
respectively. Using the branching fraction values of $\eta$ and
$\eta^\prime\to \gamma\gamma$ from the PDG~\cite{pdg}, we determine
the invisible decay rates to be $\mathcal{B}(\eta \to
\text{invisible}) < 1.0\times 10^{-4}$ and $\mathcal{B}(\eta^\prime
\to \text{invisible}) < 5.3\times 10^{-4}$ at the 90\% confidence
level.

Our limits are improved by factors of 6 and 3 compared to the
previous ones obtained at BESII~\cite{np5}, the $\eta^\prime$ limit
being almost 2 times better than the recent one from the CLEO-c
experiment~\cite{cleo-c}. The limit for $\eta\to \text{invisible}$
is smaller than a tentative estimate~\cite{np10} for the $\eta \to
\chi\chi $ decay to a pair of light dark matter particles, no such
decays, however, being expected from the virtual exchanges of a
\hbox{spin-1} $U$ boson (or dark photon) with vector couplings to
quarks. These limits constrain the decays $\eta\,(\eta')\to UU$
where each $U$ decays invisibly into neutrinos or LDM, with
branching fraction $B_{\rm inv}$. The resulting $\eta\,(\eta') $
limits on the $U$ couplings to quarks are improved by $\simeq 1.6$
and 1.3 as compared to those obtained in \cite{np1} from the BESII
limits~\cite{np5}, and now read $\sqrt{f_u^2+f_d^2} < 3 \times
10^{-2}/\sqrt{B_{\rm inv}}$ and $|f_s| < 4 \times 10^{-2}/\sqrt{B_{\rm
inv}}\,$, respectively (for $\,2m_U $ smaller than $m_\eta$  or
$m_{\eta'}$ and not too close to them), $f_u,\,f_d$ and $f_s$
denoting effective couplings of the $U$ boson to light quarks.

%%%%%%%%%%%%%%%%%%%%%%
%% End of conclusion %
%%%%%%%%%%%%%%%%%%%%%%

\begin{acknowledgements}
The BESIII collaboration thanks the staff of \nohyphens{BEPCII} and
the computing center for their hard efforts. One of the authors,
Hai-Bo Li,  thanks Pierre Fayet for illuminating suggestions. This
work is supported in part by the Ministry of Science and Technology
of China under Contract No. 2009CB825200; National Natural Science
Foundation of China (NSFC) under Contracts Nos. 10625524, 10821063,
10825524, 10835001, 10935007, 11125525, 11061140514; Joint Funds of the National
Natural Science Foundation of China under Contracts Nos. 11079008,
11179007, 11179014; the Chinese Academy of Sciences (CAS)
Large-Scale Scientific Facility Program; CAS under Contracts Nos.
KJCX2-YW-N29, KJCX2-YW-N45; 100 Talents Program of CAS; Istituto
Nazionale di Fisica Nucleare, Italy; Ministry of Development of
Turkey under Contract No. DPT2006K-120470; U. S. Department of
Energy under Contracts Nos. DE-FG02-04ER41291, DE-FG02-91ER40682,
DE-FG02-94ER40823; U.S. National Science Foundation; University of
Groningen (RuG) and the Helmholtzzentrum fuer Schwerionenforschung
GmbH (GSI), Darmstadt; WCU Program of National Research Foundation
of Korea under Contract No. R32-2008-000-10155-0.
\end{acknowledgements}

%%%%%%%%%%%%%%%%%%%%%%%%%%%%%%%%%%%%%%%%%%%%%%%%%%%%%%%%%%%%%%%%%%%%%%%%%
% BIBLIOGRAPHY
%%%%%%%%%%%%%%%%%%%%%%%%%%%%%%%%%%%%%%%%%%%%%%%%%%%%%%%%%%%%%%%%%%%%%%%%%

%\begin{thebibliography}{99}

%\bibliographystyle{h-physrev2-original}   %


\begin{thebibliography}{99}
\bibitem{pwabes2-1} It is noted that diffrent definition of Flatt$\acute{\text{e}}$
form is used in Eq. (1) in Ref.~\cite{pwabes2}. Here, $g^{2}_{1}$
and $g^{2}_{2}$ are identical to `$g_{1}$' and `$g_{2}$' in
Ref.~\cite{pwabes2}. Therefore, we notice that the unit for the
`$g_{1}$' and `$g_{2}$' in Ref.~\cite{pwabes2} should be
(GeV/c$^2$)$^2$ instead of GeV/c$^2$.
\end{thebibliography}
\end{document}